\documentclass{article}%
\usepackage{amsfonts}
\usepackage{amsmath}
\usepackage{amssymb}
\usepackage{charter}
\usepackage{graphicx}%
\providecommand{\U}[1]{\protect \rule{.1in}{.1in}}

\providecommand{\U}[1]{\protect \rule{.1in}{.1in}}
\begin{document}

\title{Entangled Husimi distribution and Complex Wavelet
transformation\thanks{{\small Corresponding author. Email address:
hlyun2008@126.com.}}}
\author{{\small Li-yun Hu}$^{1}$ {\small and Hong-yi Fan}$^{2}$\\$^{1}${\small College of Physics and Communication Electronics, Jiangxi Normal
University, Nanchang 330022, China}\\$^{2}${\small Department of Physics, Shanghai Jiao Tong University, Shanghai
200030, China}}
\maketitle

\begin{abstract}
{\small Based on the proceding Letter }[{\small Int. J. Theor. Phys. 48, 1539
(2009)}], {\small we expand the relation between wavelet transformation and
Husimi distribution function to the entangled case. We find that the optical
complex wavelet transformation can be used to study the entangled Husimi
distribution function in phase space theory of quantum optics. We prove that
the entangled Husimi distribution function of a two-mode quantum state
}$\left \vert \psi \right \rangle ${\small \ is just the modulus square of the
complex wavelet transform of }$e^{-\left \vert \eta \right \vert ^{2}/2}%
${\small \ with }$\psi \left(  \eta \right)  ${\small \ being the mother wavelet
up to a Gaussian function.}


\end{abstract}

\section{Introduction}

Studying distribution functions of density operator $\rho$ in phase space has
been a major topic in quantum statistical physics. Phase space technique has
proved very effective in various branches of physics. Among various phase
space distributions the Wigner function $F_{w}\left(  q,p\right)  $
\cite{r1,r2,r3,r4} is the most popularly used. But the Wigner distribution
function itself is not a probability distribution due to being both positive
and negative. To overcome this inconvenience, the Husimi distribution function
$F_{h}\left(  q^{\prime},p^{\prime}\right)  $ is introduced \cite{r5}, which
is defined in a manner that guarantees it to be nonnegative. On the other
hand, the optical wavelet transformations have been developed which can
overcome some shortcomings of the classical Fourier analysis and therefore has
been widely used in Fourier optics and information science since 1980s
\cite{r6,r7,r8,r9}. In the previous Letter \cite{r10}, we have employed the
optical wavelet transformation to study the Husimi distribution function for
single-mode case, and proved that the Husimi distribution function of a
quantum state $\left \vert \psi \right \rangle $\ is just the modulus square of
the wavelet transform of $e^{-x^{2}/2}$\ with $\psi \left(  x\right)  $\ being
the mother wavelet up to a Gaussian function, i.e.,
\begin{equation}
\left \langle \psi \right \vert \Delta_{h}\left(  q,p,\kappa \right)  \left \vert
\psi \right \rangle =\frac{e^{-\frac{p^{2}}{\kappa}}}{\sqrt{\pi \kappa}%
}\left \vert \int_{-\infty}^{\infty}dx\psi^{\ast}\left(  \frac{x-s}{\mu
}\right)  e^{-x^{2}/2}\right \vert ^{2}, \label{e1}%
\end{equation}
where $s=\frac{-1}{\sqrt{\kappa}}\left(  \kappa q+ip\right)  ,$ $\mu
=\sqrt{\kappa},$and $\left \langle \psi \right \vert \Delta_{h}\left(
q,p\right)  \left \vert \psi \right \rangle $ is the Husimi distribution
function,%
\begin{equation}
\left \langle \psi \right \vert \Delta_{h}\left(  q,p,\kappa \right)  \left \vert
\psi \right \rangle =2\int_{-\infty}^{\infty}dq^{\prime}dp^{\prime}F_{w}\left(
q^{\prime},p^{\prime}\right)  \exp \left[  -\kappa \left(  q^{\prime}-q\right)
^{2}-\frac{\left(  p^{\prime}-p\right)  ^{2}}{\kappa}\right]  , \label{e2}%
\end{equation}
as well as $\Delta_{h}\left(  q,p,\kappa \right)  $ is the Husimi operator,
\begin{equation}
\Delta_{h}\left(  q,p,\kappa \right)  =\frac{2\sqrt{\kappa}}{1+\kappa}%
\colon \exp \left \{  \frac{-\kappa \left(  q-Q\right)  ^{2}}{1+\kappa}%
-\frac{\left(  p-P\right)  ^{2}}{1+\kappa}\right \}  \colon, \label{e3}%
\end{equation}
here $\colon \colon$ denotes normal ordering; $Q=(a+a^{\dag})/\sqrt{2}$ and
$P=(a-a^{\dag})/(\sqrt{2}\mathtt{i})$ are the coordinate and the momentum
operator, and $a_{1},a_{1}^{\dag}$ the Bose annihilation and creation
operators, $[a,a^{\dag}]=1,a\left \vert 0\right \rangle =0$. Thus a convenient
approach for calculating various Husimi distribution functions of
miscellaneous quantum states is presented.

Recalling that in Ref.\cite{r11}, Fan and Guo have introduced the entangled
Husimi operator $\Delta_{h}\left(  \sigma,\gamma,\kappa \right)  $ which is
endowed with definite physical meaning, and find that there corresponds a
special two-mode squeezed coherent state $\left \vert \sigma,\gamma
\right \rangle _{\kappa}$ representation such that $\Delta_{h}\left(
\sigma,\gamma,\kappa \right)  $ $=$ $\left \vert \sigma,\gamma,\kappa
\right \rangle \left \langle \sigma,\gamma,\kappa \right \vert $. The entangled
Husimi operator $\Delta_{h}\left(  \sigma,\gamma,\kappa \right)  $ and the
entangled Husimi distribution $F_{h}\left(  \sigma,\gamma,\kappa \right)  $ of
quantum state $\left \vert \psi \right \rangle $ are given by%

\begin{equation}
\Delta_{h}\left(  \sigma,\gamma,\kappa \right)  =4\int d^{2}\sigma^{\prime
}d^{2}\gamma^{\prime}\Delta_{w}\left(  \sigma^{\prime},\gamma^{\prime}\right)
\exp \left \{  -\kappa \left \vert \sigma^{\prime}-\sigma \right \vert ^{2}-\frac
{1}{\kappa}\left \vert \gamma^{\prime}-\gamma \right \vert ^{2}\right \}  ,
\label{e4}%
\end{equation}
and
\begin{equation}
F_{h}\left(  \sigma,\gamma,\kappa \right)  =4\int d^{2}\sigma^{\prime}%
d^{2}\gamma^{\prime}F_{w}\left(  \sigma^{\prime},\gamma^{\prime}\right)
\exp \left \{  -\kappa \left \vert \sigma^{\prime}-\sigma \right \vert ^{2}-\frac
{1}{\kappa}\left \vert \gamma^{\prime}-\gamma \right \vert ^{2}\right \}  ,
\label{e5}%
\end{equation}
respectively, where $F_{w}\left(  \sigma^{\prime},\gamma^{\prime}\right)
=\left \langle \psi \right \vert \Delta_{w}\left(  \sigma^{\prime},\gamma
^{\prime}\right)  \left \vert \psi \right \rangle $ with $\Delta_{w}\left(
\sigma^{\prime},\gamma^{\prime}\right)  $\ being two-mode Wigner operator is
two-mode Wigner function. Thus we are naturally led to studying the entangled
Husimi distribution function from the viewpoint of wavelet transformation.

In this paper, we shall expand the relation between wavelet transformation and
Wigner-Husimi distribution function to the entangled case, that is to say, we
employ the complex wavelet transformation (CWT) to investigate the entangled
Husimi distribution function (EHDF) by bridging the relation between CWT and
EHDF. We prove that the entangled Husimi distribution function of a two-mode
quantum state $\left \vert \psi \right \rangle $\ is just the modulus square of
the complex wavelet transform of $e^{-\left \vert \eta \right \vert ^{2}/2}%
$\ with $\psi \left(  \eta \right)  $\ being the mother wavelet up to a Gaussian
function. Thus we present a convenient approach for calculating various
entangled Husimi distribution functions of miscellaneous two-mode quantum states.

\section{Complex wavelet transform and its quantum mechanical version}

In Ref.\cite{r12}, Fan and Lu have proposed the complex wavelet transform
(CWT), i.e., the CWT of a signal function $g\left(  \eta \right)  $ by $\psi$
is defined by
\begin{equation}
W_{\psi}g\left(  \mu,z\right)  =\frac{1}{\mu}\int \frac{d^{2}\eta}{\pi}g\left(
\eta \right)  \psi^{\ast}\left(  \frac{\eta-z}{\mu}\right)  , \label{e6}%
\end{equation}
whose admissibility condition for mother wavelets, $\int \frac{d^{2}\eta}{2\pi
}\psi \left(  \eta \right)  =0,$ is examined in the entangled state
representations $\left \langle \eta \right \vert $ and a family of new mother
wavelets (named the Laguerre--Gaussian wavelets) are found to match the CWT
\cite{r12}. In fact, by introducing the bipartite entangled state
representation $\left \langle \eta=\eta_{1}+\mathtt{i}\eta_{2}\right \vert
,$\cite{r13,r14}
\begin{equation}
\left \vert \eta \right \rangle =\exp \left \{  -\frac{1}{2}\left \vert
\eta \right \vert ^{2}+\eta a_{1}^{\dagger}-\eta^{\ast}a_{2}^{\dagger}%
+a_{1}^{\dagger}a_{2}^{\dagger}\right \}  \left \vert 00\right \rangle ,
\label{e7}%
\end{equation}
which is the common eigenvector of relative coordinate $Q_{1}-Q_{2}$ and the
total momentum $P_{1}+P_{2}$,
\begin{equation}
\left(  Q_{1}-Q_{2}\right)  \left \vert \eta \right \rangle =\sqrt{2}\eta
_{1}\left \vert \eta \right \rangle ,\text{ }\left(  P_{1}+P_{2}\right)
\left \vert \eta \right \rangle =\sqrt{2}\eta_{2}\left \vert \eta \right \rangle ,
\label{3.12}%
\end{equation}
where $Q_{j}$ and $P_{j}$ are the coordinate and the momentum operator,
related to the Bose operators $(a_{j},a_{j}^{\dag}),[a_{i},a_{j}^{\dag
}]=\delta_{ij}$ by $Q_{j}=(a_{j}+a_{j}^{\dagger})/\sqrt{2}$\ and\ $P_{j}%
=(a-a^{\dagger})/(\sqrt{2}\mathtt{i})$ ($j=1,2$), we can treat (\ref{e5}) from
the quantum mechanically,
\begin{equation}
W_{\psi}g\left(  \mu,z\right)  =\frac{1}{\mu}\int \frac{d^{2}\eta}{\pi
}\left \langle \psi \right \vert \left.  \frac{\eta-z}{\mu}\right \rangle
\left \langle \eta \right \vert \left.  g\right \rangle =\left \langle
\psi \right \vert U_{2}\left(  \mu,z\right)  \left \vert g\right \rangle ,
\label{e8}%
\end{equation}
where $z=z_{1}+iz_{2}\in C,$ $0<\mu \in R,$ $g\left(  \eta \right)
\equiv \left \langle \eta \right \vert \left.  g\right \rangle ,\ $and $\psi \left(
\eta \right)  =\left \langle \eta \right \vert \left.  \psi \right \rangle $ are the
wavefunction of state vector $\left \vert g\right \rangle $ and the mother
wavelet state vector $\left \vert \psi \right \rangle $ in $\left \langle
\eta \right \vert $ representation, respectively, and
\begin{equation}
U_{2}\left(  \mu,z\right)  \equiv \frac{1}{\mu}\int \frac{d^{2}\eta}{\pi
}\left \vert \frac{\eta-z}{\mu}\right \rangle \left \langle \eta \right \vert ,\;
\mu=e^{\lambda}, \label{e9}%
\end{equation}
is the two-mode squeezing-displacing operator \cite{r15,r16,r17}. Noticing
that the two-mode squeezing operator has its natural expression in
$\left \langle \eta \right \vert $ representation \cite{r14},
\begin{equation}
S_{2}\left(  \mu \right)  =\exp \left[  \left(  a_{1}^{\dagger}a_{2}^{\dagger
}-a_{1}a_{2}\right)  \ln \mu \right]  =\frac{1}{\mu}\int \frac{d^{2}\eta}{\pi
}\left \vert \frac{\eta}{\mu}\right \rangle \left \langle \eta \right \vert ,
\label{e10}%
\end{equation}
which is differerent from the product of two single-mode squeezing (dilation)
operators, and the two-mode squeezed state is simultaneously an entangled
state, thus we can put Eq.(\ref{e9}) into the following form,%
\begin{equation}
U_{2}\left(  \mu,z\right)  =S_{2}\left(  \mu \right)  \mathfrak{D}\left(
z\right)  , \label{e11}%
\end{equation}
where $\mathfrak{D}\left(  z\right)  $ is a two-mode displacement operator,
$\mathfrak{D}\left(  z\right)  \left \vert \eta \right \rangle =\left \vert
\eta-z\right \rangle $ and
\begin{align}
\mathfrak{D}\left(  z\right)   &  =\int \frac{d^{2}\eta}{\pi}\left \vert
\eta-z\right \rangle \left \langle \eta \right \vert \nonumber \\
&  =\exp \left[  iz_{1}\frac{P_{1}-P_{2}}{\sqrt{2}}-iz_{2}\frac{Q_{1}+Q_{2}%
}{\sqrt{2}}\right] \nonumber \\
&  =D_{1}\left(  -z/2\right)  D_{2}\left(  z^{\ast}/2\right)  . \label{e12}%
\end{align}
It the follows the quantum mechanical version of CWT is%
\begin{equation}
W_{\psi}g\left(  \mu,\zeta \right)  =\left \langle \psi \right \vert S_{2}\left(
\mu \right)  \mathfrak{D}\left(  z\right)  \left \vert g\right \rangle
=\left \langle \psi \right \vert S_{2}\left(  \mu \right)  D_{1}\left(
-z/2\right)  D_{2}\left(  z^{\ast}/2\right)  \left \vert g\right \rangle .
\label{e13}%
\end{equation}
Eq.(\ref{e13}) indicates that the 2D CWT can be put into a matrix element in
the $\left \langle \eta \right \vert $ representation of the two-mode displacing
and the two-mode squeezing operators in Eq.(\ref{e10}) between the mother
wavelet state vector $\left \vert \psi \right \rangle $ and the state vector
$\left \vert g\right \rangle $ to be transformed. Thus the CWT differs from the
direct product of two 1-dimensional wavelet transformations.

Once the state vector $\left \langle \psi \right \vert $ corresponding to mother
wavelet is known, for any state $\left \vert g\right \rangle $ the matrix
element $\left \langle \psi \right \vert U_{2}\left(  \mu,z\right)  \left \vert
g\right \rangle $ is just the wavelet transform of $g(\eta)$ with respect to
$\left \langle \psi \right \vert .$ Therefore, various quantum optical field
states can then be analyzed by their wavelet transforms.

\section{Relation between CWT and EHDF}

In the following we shall show that the entangled Husimi distribution function
(EHDF) of a quantum state $\left \vert \psi \right \rangle $ can be obtained by
making a complex wavelet transform of the Gaussian function $e^{-\left \vert
\eta \right \vert ^{2}/2},$ i.e.,
\begin{equation}
\left \langle \psi \right \vert \Delta_{h}\left(  \sigma,\gamma,\kappa \right)
\left \vert \psi \right \rangle =e^{-\frac{1}{\kappa}\left \vert \gamma \right \vert
^{2}}\left \vert \int \frac{d^{2}\eta}{\sqrt{\kappa}\pi}e^{-\left \vert
\eta \right \vert ^{2}/2}\psi^{\ast}\left(  \frac{\eta-z}{\sqrt{\kappa}}\right)
\right \vert ^{2}, \label{e14}%
\end{equation}
where $\mu=e^{\lambda}=\sqrt{\kappa},$ $z=z_{1}+iz_{2},$ and
\begin{align}
z_{1}  &  =\frac{\cosh \lambda}{1+\kappa}\left[  \gamma^{\ast}-\gamma
-\kappa \left(  \sigma^{\ast}+\sigma \right)  \right]  ,\label{e15}\\
z_{2}  &  =\frac{i\cosh \lambda}{1+\kappa}\left[  \gamma+\gamma^{\ast}%
+\kappa \left(  \sigma-\sigma^{\ast}\right)  \right]  , \label{e16}%
\end{align}
and $\left \langle \psi \right \vert \Delta_{h}\left(  \sigma,\gamma
,\kappa \right)  \left \vert \psi \right \rangle $ is the Husimi distribution
function as well as $\Delta_{h}\left(  \sigma,\gamma,\kappa \right)  $ is the
Husimi operator,
\begin{align}
\Delta_{h}\left(  \sigma,\gamma,\kappa \right)   &  =\frac{4\kappa}{\left(
1+\kappa \right)  ^{2}}\colon \exp \left \{  -\frac{\left(  a_{1}+a_{2}^{\dag
}-\gamma \right)  \left(  a_{1}^{\dag}+a_{2}-\gamma^{\ast}\right)  }{1+\kappa
}\right. \nonumber \\
&  -\left.  \frac{\kappa \left(  a_{1}-a_{2}^{\dag}-\sigma \right)  \left(
a_{1}^{\dag}-a_{2}-\sigma^{\ast}\right)  }{1+\kappa}\right \}  \colon
\label{e17}%
\end{align}
here $\colon \colon$ denotes normal ordering of operators.

\textbf{Proof of Eq.(\ref{e14}).}

When the transformed $\left \vert g\right \rangle =\left \vert 00\right \rangle $
(the two-mode vacuum state), noticing that $\left \langle \eta \right.
\left \vert 00\right \rangle =e^{-\left \vert \eta \right \vert ^{2}/2},$ thus we
can express Eq.(\ref{e8}) as%
\begin{equation}
\frac{1}{\mu}\int \frac{d^{2}\eta}{\pi}e^{-\left \vert \eta \right \vert ^{2}%
/2}\psi^{\ast}\left(  \frac{\eta-z}{\mu}\right)  =\left \langle \psi \right \vert
U_{2}\left(  \mu,z\right)  \left \vert 00\right \rangle . \label{e18}%
\end{equation}
To combine the CWTs with transforms of quantum states more tightly and
clearly, using the technique of integration within an ordered product (IWOP)
\cite{r18,r19,r20,r21} of operators, we can directly perform the integral in
Eq.(\ref{e9}) \cite{r22}%
\begin{align}
U_{2}\left(  \mu,z\right)   &  =\frac{1}{\mu}\int \frac{d^{2}\eta}{\pi}%
\colon \exp \left \{  -\frac{\mu^{2}+1}{2\mu^{2}}|\eta|^{2}+\frac{\eta z^{\ast
}+z\eta^{\ast}}{2\mu^{2}}+\frac{\eta-z}{\mu}a_{1}^{\dagger}\right. \nonumber \\
&  \left.  -\frac{\eta^{\ast}-z^{\ast}}{\mu}a_{2}^{\dagger}+a_{1}^{\dagger
}a_{2}^{\dagger}+\eta^{\ast}a_{1}-\eta a_{2}+a_{1}a_{2}-a_{1}^{\dagger}%
a_{1}-a_{2}^{\dagger}a_{2}-\frac{\left \vert z\right \vert ^{2}}{2\mu^{2}%
}\right \}  \colon \nonumber \\
&  =\operatorname{sech}\lambda \exp \left[  -\frac{1}{2\left(  1+\mu^{2}\right)
}\left \vert z\right \vert ^{2}+a_{1}^{\dagger}a_{2}^{\dagger}\tanh \lambda
+\frac{1}{2}\left(  z^{\ast}a_{2}^{\dagger}-za_{1}^{\dagger}\right)
\operatorname{sech}\lambda \right] \nonumber \\
&  \times \exp \left[  \left(  a_{1}^{\dagger}a_{1}+a_{2}^{\dagger}a_{2}\right)
\ln \operatorname{sech}\lambda \right]  \exp \left(  \frac{z^{\ast}a_{1}-za_{2}%
}{1+\mu^{2}}-a_{1}a_{2}\tanh \lambda \right)  . \label{e19}%
\end{align}
where we have set $\mu=e^{\lambda}$, $\operatorname{sech}\lambda=\frac{2\mu
}{1+\mu^{2}}$, $\tanh \lambda=\frac{\mu^{2}-1}{\mu^{2}+1}$, and we have used
the operator identity $e^{ga^{\dagger}a}=\colon \exp \left[  \left(
e^{g}-1\right)  a^{\dagger}a\right]  \colon$. In particular, when $z=0,$ it
reduces to the usual normally ordered two-mode squeezing operator
$S_{2}\left(  \mu \right)  $. From Eq.(\ref{e19}) it then follows that
\begin{align}
U_{2}\left(  \mu,z\right)  \left \vert 00\right \rangle  &  =\operatorname{sech}%
\lambda \exp \left \{  -\frac{\left(  z_{1}-iz_{2}\right)  \left(  z_{1}%
+iz_{2}\right)  }{2\left(  1+\mu^{2}\right)  }+a_{1}^{\dagger}a_{2}^{\dagger
}\tanh \lambda \right. \nonumber \\
&  \left.  +\frac{1}{2}\left[  \left(  z_{1}-iz_{2}\right)  a_{2}^{\dagger
}-\left(  z_{1}+iz_{2}\right)  a_{1}^{\dagger}\right]  \operatorname{sech}%
\lambda \right \}  \left \vert 00\right \rangle . \label{e20}%
\end{align}
Substituting Eqs.(\ref{e15}), (\ref{e16}) and $\tanh \lambda=\frac{\kappa
-1}{\kappa+1},$ $\cosh \lambda=\frac{1+\kappa}{2\sqrt{\kappa}}$ into
Eq.(\ref{e20}) yields%
\begin{align}
&  e^{-\frac{1}{2\kappa}\left \vert \gamma \right \vert ^{2}-\frac{\sigma
\gamma^{\ast}-\gamma \sigma^{\ast}}{2\left(  \kappa+1\right)  }}U_{2}\left(
\mu,z_{1},z_{2}\right)  \left \vert 00\right \rangle \nonumber \\
&  =\frac{2\sqrt{\kappa}}{1+\kappa}\exp \left \{  -\frac{\left \vert
\gamma \right \vert ^{2}+\kappa \left \vert \sigma \right \vert ^{2}}{2\left(
\kappa+1\right)  }+\frac{\kappa \sigma+\gamma}{1+\kappa}a_{1}^{\dagger}%
+\frac{\gamma^{\ast}-\kappa \sigma^{\ast}}{1+\kappa}a_{2}^{\dagger}%
+a_{1}^{\dagger}a_{2}^{\dagger}\frac{\kappa-1}{\kappa+1}\right \}
\allowbreak \left \vert 00\right \rangle \left.  \equiv \right.  \left \vert
\sigma,\gamma \right \rangle _{\kappa}, \label{e21}%
\end{align}
then the CWT of Eq.(\ref{e18}) can be further expressed as%
\begin{equation}
e^{-\frac{1}{2\kappa}\left \vert \gamma \right \vert ^{2}-\frac{\sigma
\gamma^{\ast}-\gamma \sigma^{\ast}}{2\left(  \kappa+1\right)  }}\int \frac
{d^{2}\eta}{\mu \pi}e^{-\left \vert \eta \right \vert ^{2}/2}\psi^{\ast}\left(
\frac{\eta-z_{1}-iz_{2}}{\mu}\right)  =\left \langle \psi \right.  \left \vert
\sigma,\gamma \right \rangle _{\kappa}. \label{e22}%
\end{equation}

Using normally ordered form of the vacuum state projector $\left \vert
00\right \rangle \left \langle 00\right \vert =\colon e^{-a_{1}^{\dagger}%
a_{1}-a_{2}^{\dagger}a_{2}}\colon,$and the IWOP method as well as
Eq.(\ref{e21}) we have%
\begin{align}
\left \vert \sigma,\gamma \right \rangle _{\kappa \kappa}\left \langle
\sigma,\gamma \right \vert  &  =\frac{4\kappa}{\left(  1+\kappa \right)  ^{2}%
}\colon \exp \left[  -\frac{\left \vert \gamma \right \vert ^{2}+\kappa \left \vert
\sigma \right \vert ^{2}}{\kappa+1}+\frac{\kappa \sigma+\gamma}{1+\kappa}%
a_{1}^{\dagger}+\frac{\gamma^{\ast}-\kappa \sigma^{\ast}}{1+\kappa}%
a_{2}^{\dagger}\right. \nonumber \\
&  \left.  +\frac{\kappa \sigma^{\ast}+\gamma^{\ast}}{1+\kappa}a_{1}%
+\frac{\gamma-\kappa \sigma}{1+\kappa}a_{2}+\frac{\kappa-1}{\kappa+1}\left(
a_{1}^{\dagger}a_{2}^{\dagger}+a_{1}a_{2}\right)  -a_{1}^{\dagger}a_{1}%
-a_{2}^{\dagger}a_{2}\right]  \colon \nonumber \\
&  =\frac{4\kappa}{\left(  1+\kappa \right)  ^{2}}\colon \exp \left \{
-\frac{\left(  a_{1}+a_{2}^{\dag}-\gamma \right)  \left(  a_{1}^{\dag}%
+a_{2}-\gamma^{\ast}\right)  }{1+\kappa}\right. \nonumber \\
&  -\left.  \frac{\kappa \left(  a_{1}-a_{2}^{\dag}-\sigma \right)  \left(
a_{1}^{\dag}-a_{2}-\sigma^{\ast}\right)  }{1+\kappa}\right \}  \colon \left.
=\right.  \Delta_{h}\left(  \sigma,\gamma,\kappa \right)  . \label{e23}%
\end{align}
Now we explain why $\Delta_{h}\left(  \sigma,\gamma,\kappa \right)  $ is the
entangled Husimi operator. Using the formula for converting an operator $A$
into its Weyl ordering form \cite{r23}%
\begin{equation}
A=4\int \frac{d^{2}\alpha d^{2}\beta}{\pi^{2}}\left \langle -\alpha
,-\beta \right \vert A\left \vert \alpha,\beta \right \rangle
\genfrac{}{}{0pt}{}{:}{:}%
\exp \{2\left(  \alpha^{\ast}a_{1}-a_{1}^{\dagger}\alpha+\beta^{\ast}%
a_{2}-a_{2}^{\dagger}\beta+a_{1}^{\dagger}a_{1}+a_{2}^{\dagger}a_{2}\right)
\}%
\genfrac{}{}{0pt}{}{:}{:}%
, \label{e24}%
\end{equation}
where the symbol $%
\genfrac{}{}{0pt}{}{:}{:}%
\genfrac{}{}{0pt}{}{:}{:}%
$ denotes the Weyl ordering, $\left \vert \beta \right \rangle $ is the usual
coherent state, substituting\ Eq.(\ref{e23}) into Eq.(\ref{e24}) and
performing the integration by virtue of the technique of integration within a
Weyl ordered product of operators, we obtain%
\begin{align}
\left \vert \sigma,\gamma \right \rangle _{\kappa \kappa}\left \langle
\sigma,\gamma \right \vert  &  =\frac{16\kappa}{\left(  1+\kappa \right)  ^{2}%
}\int \frac{d^{2}\alpha d^{2}\beta}{\pi^{2}}\left \langle -\alpha,-\beta
\right \vert \colon \exp \left \{  -\frac{\left(  a_{1}+a_{2}^{\dag}%
-\gamma \right)  \left(  a_{1}^{\dag}+a_{2}-\gamma^{\ast}\right)  }{1+\kappa
}\right. \nonumber \\
&  \left.  -\frac{\kappa \left(  a_{1}-a_{2}^{\dag}-\sigma \right)  \left(
a_{1}^{\dag}-a_{2}-\sigma^{\ast}\right)  }{1+\kappa}\right \}  \colon \left \vert
\alpha,\beta \right \rangle \nonumber \\
&  \times%
\genfrac{}{}{0pt}{}{:}{:}%
\exp \{2\left(  \alpha^{\ast}a_{1}-a_{1}^{\dagger}\alpha+\beta^{\ast}%
a_{2}-a_{2}^{\dagger}\beta+a_{1}^{\dagger}a_{1}+a_{2}^{\dagger}a_{2}\right)
\}%
\genfrac{}{}{0pt}{}{:}{:}%
\nonumber \\
&  =4%
\genfrac{}{}{0pt}{}{:}{:}%
\exp \left \{  -\kappa \left(  a_{1}-a_{2}^{\dag}-\sigma \right)  \left(
a_{1}^{\dag}-a_{2}-\sigma^{\ast}\right)  -\frac{1}{\kappa}\left(  a_{1}%
+a_{2}^{\dag}-\gamma \right)  \left(  a_{1}^{\dag}+a_{2}-\gamma^{\ast}\right)
\right \}
\genfrac{}{}{0pt}{}{:}{:}%
, \label{e25}%
\end{align}
where we have used the integral formula%
\begin{equation}
\int \frac{d^{2}z}{\pi}\exp \left(  \zeta \left \vert z\right \vert ^{2}+\xi z+\eta
z^{\ast}\right)  =-\frac{1}{\zeta}e^{-\frac{\xi \eta}{\zeta}},\text{Re}\left(
\zeta \right)  <0. \label{e26}%
\end{equation}
This is the Weyl ordering form of $\left \vert \sigma,\gamma \right \rangle
_{\kappa \kappa}\left \langle \sigma,\gamma \right \vert .$ Then according to Weyl
quantization scheme \cite{r24,r25} we know the Weyl ordering form of two-mode
Wigner operator is given by%
\begin{equation}
\Delta_{w}\left(  \sigma,\gamma \right)  =%
\genfrac{}{}{0pt}{}{:}{:}%
\delta \left(  a_{1}-a_{2}^{\dag}-\sigma \right)  \delta \left(  a_{1}^{\dag
}-a_{2}-\sigma^{\ast}\right)  \delta \left(  a_{1}+a_{2}^{\dag}-\gamma \right)
\delta \left(  a_{1}^{\dag}+a_{2}-\gamma^{\ast}\right)
\genfrac{}{}{0pt}{}{:}{:}%
, \label{e27}%
\end{equation}
thus the classical corresponding function of a Weyl ordered operator is
obtained by just replacing $a_{1}-a_{2}^{\dag}\rightarrow \sigma^{\prime}%
,a_{1}+a_{2}^{\dag}\rightarrow \gamma^{\prime},$ i.e.,%
\begin{align}
&  4%
\genfrac{}{}{0pt}{}{:}{:}%
\exp \left \{  -\kappa \left(  a_{1}-a_{2}^{\dag}-\sigma \right)  \left(
a_{1}^{\dag}-a_{2}-\sigma^{\ast}\right)  -\frac{1}{\kappa}\left(  a_{1}%
+a_{2}^{\dag}-\gamma \right)  \left(  a_{1}^{\dag}+a_{2}-\gamma^{\ast}\right)
\right \}
\genfrac{}{}{0pt}{}{:}{:}%
\nonumber \\
&  \rightarrow4\exp \left \{  -\kappa \left \vert \sigma^{\prime}-\sigma
\right \vert ^{2}-\frac{1}{\kappa}\left \vert \gamma^{\prime}-\gamma \right \vert
^{2}\right \}  , \label{e28}%
\end{align}
and in this case the Weyl rule is expressed as%
\begin{align}
\left \vert \sigma,\gamma \right \rangle _{\kappa \kappa}\left \langle
\sigma,\gamma \right \vert  &  =4\int d^{2}\sigma^{\prime}d^{2}\gamma^{\prime}%
\genfrac{}{}{0pt}{}{:}{:}%
\delta \left(  a_{1}-a_{2}^{\dag}-\sigma \right)  \delta \left(  a_{1}^{\dag
}-a_{2}-\sigma^{\ast}\right)  \delta \left(  a_{1}+a_{2}^{\dag}-\gamma \right)
\nonumber \\
&  \times \delta \left(  a_{1}^{\dag}+a_{2}-\gamma^{\ast}\right)
\genfrac{}{}{0pt}{}{:}{:}%
\exp \left \{  -\kappa \left \vert \sigma^{\prime}-\sigma \right \vert ^{2}-\frac
{1}{\kappa}\left \vert \gamma^{\prime}-\gamma \right \vert ^{2}\right \}
\nonumber \\
&  =4\int d^{2}\sigma^{\prime}d^{2}\gamma^{\prime}\Delta_{w}\left(
\sigma^{\prime},\gamma^{\prime}\right)  \exp \left \{  -\kappa \left \vert
\sigma^{\prime}-\sigma \right \vert ^{2}-\frac{1}{\kappa}\left \vert
\gamma^{\prime}-\gamma \right \vert ^{2}\right \}  . \label{e29}%
\end{align}
In reference to Eq.(\ref{e5}) in which the relation between the entangled
Husimi function and the two-mode Wigner function is shown, we know that the
right-hand side of Eq. (\ref{e29}) should be just the entangled Husimi
operator, i.e.
\begin{equation}
\left \vert \sigma,\gamma \right \rangle _{\kappa \kappa}\left \langle
\sigma,\gamma \right \vert =4\int d^{2}\sigma^{\prime}d^{2}\gamma^{\prime}%
\Delta_{w}\left(  \sigma^{\prime},\gamma^{\prime}\right)  \exp \left \{
-\kappa \left \vert \sigma^{\prime}-\sigma \right \vert ^{2}-\frac{1}{\kappa
}\left \vert \gamma^{\prime}-\gamma \right \vert ^{2}\right \}  =\Delta_{h}\left(
\sigma,\gamma,\kappa \right)  , \label{e30}%
\end{equation}
thus Eq. (\ref{e14}) is proved by combining Eqs.(\ref{e30}) and (\ref{e22}).

Motivated by the proceding Letter \cite{r10}, we have futher expanded the
relation between wavelet transformation and Wigner-Husimi distribution
function to the entangled case. That is to say, we prove that the entangled
Husimi distribution function of a two-mode quantum state $\left \vert
\psi \right \rangle $\ is just the modulus square of the complex wavelet
transform of $e^{-\left \vert \eta \right \vert ^{2}/2}$\ with $\psi \left(
\eta \right)  $\ being the mother wavelet up to a Gaussian function, i.e.,
$\left \langle \psi \right \vert \Delta_{h}\left(  \sigma,\gamma,\kappa \right)
\left \vert \psi \right \rangle =e^{-\frac{1}{\kappa}\left \vert \gamma \right \vert
^{2}}\left \vert \int \frac{d^{2}\eta}{\sqrt{\kappa}\pi}e^{-\left \vert
\eta \right \vert ^{2}/2}\psi^{\ast}\left(  \left(  \eta-z\right)  /\sqrt
{\kappa}\right)  \right \vert ^{2}$. Thus we have a convenient approach for
calculating various entangled Husimi distribution functions of miscellaneous
quantum states. For more discussion about the wavelet transformation in the
context of quantum optics, we refer to Refs.\cite{r26,r27}.

\textbf{Acknowledgement} Work supported by the National Natural Science
Foundation of China under grants. 10775097 and 10874174, and the Research
Foundation of the Education Department of Jiangxi Province.

\textbf{Appendix}

We can check Eq.(\ref{e30}) by the following way.

Using the normally ordered form of the two-mode Wigner operator \cite{r11}%
\begin{equation}
\Delta_{w}\left(  \sigma,\gamma \right)  =\frac{1}{\pi^{2}}\colon \exp \left \{
-\left(  a_{1}-a_{2}^{\dag}-\sigma \right)  \left(  a_{1}^{\dag}-a_{2}%
-\sigma^{\ast}\right)  -\left(  a_{1}+a_{2}^{\dag}-\gamma \right)  \left(
a_{1}^{\dag}+a_{2}-\gamma^{\ast}\right)  \right \}  \colon, \tag{A1}%
\end{equation}
we can further perform the integration in Eq.(\ref{e4}) and see%
\begin{align}
\Delta_{h}\left(  \sigma,\gamma,\kappa \right)   &  =4\int \frac{d^{2}%
\sigma^{\prime}d^{2}\gamma^{\prime}}{\pi^{2}}\exp \left \{  -\kappa \left \vert
\sigma^{\prime}-\sigma \right \vert ^{2}-\frac{1}{\kappa}\left \vert
\gamma^{\prime}-\gamma \right \vert ^{2}\right \} \nonumber \\
&  \times \colon \exp \left \{  -\left(  a_{1}-a_{2}^{\dag}-\sigma \right)  \left(
a_{1}^{\dag}-a_{2}-\sigma^{\ast}\right)  -\left(  a_{1}+a_{2}^{\dag}%
-\gamma \right)  \left(  a_{1}^{\dag}+a_{2}-\gamma^{\ast}\right)  \right \}
\colon \nonumber \\
&  =\frac{4\kappa}{\left(  1+\kappa \right)  ^{2}}\colon \exp \left \{
-\frac{\left(  a_{1}+a_{2}^{\dag}-\gamma \right)  \left(  a_{1}^{\dag}%
+a_{2}-\gamma^{\ast}\right)  }{1+\kappa}-\frac{\kappa \left(  a_{1}-a_{2}%
^{\dag}-\sigma \right)  \left(  a_{1}^{\dag}-a_{2}-\sigma^{\ast}\right)
}{1+\kappa}\right \}  \colon \nonumber \\
&  =\text{Eq.(\ref{e23})}\left.  =\right.  \Delta_{h}\left(  \sigma
,\gamma,\kappa \right)  , \tag{A2}%
\end{align}
which is the confirmation of Eq. (\ref{e30}).

\end{document}